\documentstyle[psfig, epsfig, 12pt]{article} 

\def\mypagenumber{1}
\def\myend{\end{document}}

\newcommand{\plb}[3]{{{\it Phys.~Lett.}~{\bf B#1} (#3) #2}}

\newcommand{\prd}[3]{{{\it Phys.~Rev.}~{\bf D#1} (#3) #2}}

\newcommand{\prl}[3]{{{\it Phys.~Rev.~Lett.}~{\bf #1} (#3) #2}}

\newcommand{\hepth}[1]{{\tt hep-th/#1}}

\def\lim{\mbox{{\bf L}} }

\newpage 
\normalsize

\newcounter{sxn}

\newcounter{axn}

\date{}

\newdimen\mybaselineskip
\newcommand{\beq}{\begin{equation}}
\newcommand{\eeq}{\end{equation}}
\newcommand{\bea}{\begin{eqnarray}}
\newcommand{\eea}{\end{eqnarray}}
\newcommand{\ba}{\begin{eqnarray}}
\newcommand{\ea}{\end{eqnarray}}
\newcommand{\bpic}{\begin{picture}}
\newcommand{\epic}{\end{picture}}

\def\la{\raise.16ex\hbox{$\langle$} \, }
\def\ra{\, \raise.16ex\hbox{$\rangle$} }

\def\psibar{ \psi \kern-.65em\raise.6em\hbox{$-$} }
\def\mbar{ m \kern-.78em\raise.4em\hbox{$-$}\lower.4em\hbox{} }

\def\n@space{\nulldelimiterspace=0pt \mathsurround=0pt }
\def\huge#1{{\hbox{$\left#1\vbox to 20.5pt{}\right.\n@space$}}}

\def\myskip{\noalign{\kern 8pt}}
\def\myeqspace{\noalign{\kern 10pt}}

\def\boxit#1{$\vcenter{\hrule\hbox{\vrule\kern3pt
    \vbox{\kern3pt\hbox{#1}\kern3pt}\kern3pt\vrule}\hrule}$}
\def\bigbox#1{$\vcenter{\hrule\hbox{\vrule\kern5pt
     \vbox{\kern5pt\hbox{#1}\kern5pt}\kern5pt\vrule}\hrule}$}

\def\ignore#1{{}}

\begin{document}

\bibliographystyle{unsrt}
\footskip 1.0cm

\thispagestyle{empty}
\setcounter{page}{\mypagenumber}

             
\begin{flushright}{CERN-TH/2001-088
MIT-CTP-3095 \\
hep-th/0103235
}

\end{flushright}

\vspace{1.cm}
\begin{center}
{\Large \bf {Hagedorn transition, vortices and $D0$
branes:\\Lessons from $2+1$ confining strings.}}
\vspace {.5cm}
{\Large  \bf {}}\
\

{\bf 
Ian I. Kogan$^a${\footnote{kogan@thphys.ox.ac.uk}},
Alex Kovner$^{b,c}${\footnote{a.kovner@plymouth.ac.uk}},
Martin Schvellinger$^{a,d}${\footnote{martin@thphys.ox.ac.uk}}
}\\
\vspace{.5cm}
$^a${\it Theoretical Physics, Department of Physics,
University of Oxford, 1 Keble Road, Oxford, OX1 3NP, UK}
\\
\vspace{0.2 cm}
$^b${\it Theory Division, CERN, CH-1211, Geneva 23, Switzerland}\\
\vskip 0.2 cm
$^c${\it Department of Mathematics and Statistics, 
University of Plymouth,\\
Plymouth PL4 8AA, UK}\\
$^d${\it Center for Theoretical Physics,
Massachusetts Institute of Technology, \\
77 Massachusetts Avenue, Cambridge MA 02139, USA}\\
\end{center}

\vspace*{.5cm}


\begin{abstract}
\baselineskip=15pt
We study the  behaviour of Polyakov confining string in the
Georgi-Glashow model in three dimensions near confining-deconfining 
phase transition described in \cite{KOVNER00}. 
In the string language, the transition mechanism is the decay of the 
confining string into $D0$ branes (charged $W^\pm$ bosons of the
Georgi-Glashow
model). In the world-sheet picture the world-lines of heavy 
$D0$ branes at finite temperature are represented as 
world-sheet vortices of a certain type, and the transition corresponds
to the condensation of these vortices.
We also show that the ``would be'' Hagedorn transition in the confining string
(which is not realized in our model) 
corresponds to the monopole binding transition in the field
theoretical language. The fact that the decay into $D0$ branes occurs
at lower than the Hagedorn temperature is understood as the 
consequence of the
large thickness of the confining string and finite mass of the $D0$ branes. 
\end{abstract}
\vfill
CERN-TH/2001-088


 
\newpage



\normalsize
\baselineskip=20pt plus 1pt minus 1pt
\parindent=25pt

\newpage

\section{Introduction}

What happens to strings at very high temperatures is one of the Big
 questions of string theory. In the early days of  dual resonance models and
 hadronic bootstrap the exponential growth in the one-particle density of 
states   
\beq 
\label{hag}
\rho(m) \sim  m^a \,\,e^{bm}
\eeq
led to the famous Hagedorn transition
\cite{HAGEDORN}.
This type of spectrum first arose in the context of statistical
bootstrap models~\cite{HAGEDORN,FVHW,FC} and, for
 hadrons, such behaviour indicates that they are composed of more 
fundamental constituents \cite{cab}. In  
fundamental string theories we find  the same kind of 
spectrum (see for example  \cite{BOOKS}), and a search for hints to the
existence of `string constituents' is of great interest. 
What lies beyond the Hagedorn temperature? Is this temperature limiting or 
is there a high-temperature phase which reveals the fundamental degrees of 
freedom?  And is it true that for all types of string theories there is 
the same universal physics  or different classes may have totally different 
high-temperature behaviour?

A lot of effort has been invested into
the study of the Hagedorn transition in the 
critical (super)strings. 
There is enormous literature on this subject, some references (but by no means
all) can  be found, for example in \cite{ABKR}.
 For weakly coupled critical (super)strings  the Hagedorn transition  can be
described as  Berezinsky-Kosterlitz-Thouless  (BKT)
\cite{BEREZINSKY,KOSTERLITZ} transition on a world-sheet
\cite{KOGAN87,SATHIAPALAN87, ATICK, AK}.  It 
is due to the "condensation" of the world sheet vortices. 
It has been also suggested that
the transition in some cases  may
actually be first order \cite{ATICK} (see
also discussion in \cite{AK})\footnote{ 
 In the recent paper \cite{noncomm} it was shown that 
in non-commutative open string theory  which is decoupled from gravity,
in no more than five space-time dimensions the Hagedorn transition 
is second order.}. Above the Hagedorn temperature the vortices 
populate the world-sheet and  we have a new phase. From the target space 
point of view we are talking about tachyonic instabilities for non zero winding
modes in the imaginary time direction (thermal winding 
modes)\footnote{In some cases  extra symmetry
can protect the theory from tachyonic instabilities.  For example it was
shown in \cite{N4} that the theory with ${\cal{N}}=4$ supersymmetry
admits BPS solutions that do not suffer from Hagedorn-type
instabilities. However for generic string theory these instabilities
must happen.}.

Although much have been understood, many aspects
of hot string theory remain mysterious.
For example, it is thought that for critical
(super)strings the whole idea of using  canonical Gibbs ensemble
 may be not justified, since the canonical and the micro-canonical ensembles 
are found to be inequivalent \cite{FC}, \cite{deo}. Besides 
the whole notion of temperature and canonical ensemble is not well
defined in the presence of gravity (see \cite{ATICK} and references
therein)  which necessarily exists in theories of critical strings.
In the more general case of interacting strings we do not know with certainty
what the fundamental degrees of freedom are\footnote{M,F,S theories, Matrix
Models, etc.  give us a lot of new exciting hints,
but we still do not have a solid theoretical
concept.} and thus what is their role at high temperature.
Recently it was suggested in the framework of the 
Matrix Model description of the Hagedorn transition, that
the fundamental string decays into $D0$ branes \cite{SATH} (for
references on thermodynamics in Matrix String theory see \cite{MATRIX} 
and references therein). Thus it could be that $D0$ branes are the fundamental
degrees of freedom in the hot phase.

Since non Abelian confining gauge theories are strongly linked to strings,
we may hope to glean some insight about the high temperature behaviour 
of strings from the study of the deconfined phase of gauge theories.
When QCD emerged as the fundamental theory of strong interactions,
it was suggested that there is a 
deconfining phase transition \cite{POLZN}  during which the
symmetry of the centre of the gauge group in electric representation is 
broken.  In the  dual magnetic picture this corresponds to the restoration 
of the true global magnetic symmetry $Z_N$ 
(for $SU(N)$ gauge theory)\cite{thooft1} -  for more details 
see \cite{KOVNERZN}. The lore is that  these two transitions are the same. 
The confining phase of QCD should be describable by some  string theory. 
 At least in the large $N$ limit we hope it is a weakly coupled string
with coupling of order $1/N$  \cite{THOOFT}.  
Above the deconfining transition  the
string description does not make sense. This must mean that the 
string itself undergoes some kind of a phase transition.  This  picture is 
very attractive. If this is the case, 
we are definitely not looking for a grey cat in a dark room, 
since  the high temperature phase is described in terms of 
quarks and gluons and we
know the fundamental Lagrangian which determines their properties. 
The problem is that we still have no idea what string theory 
we are dealing with (recent results based on AdS/CFT 
duality \cite{ADSCFT} may turn out eventually to be useful in this respect)  
for the simple reason: so far  there is no consistent theory of QCD
confinement. It may in fact be quite an unusual string theory. For example,
the analysis of the 
high-temperature partition function  of a chromoelectric
flux tube in the large $N$ limit suggests  that 
the effective string at short distances 
has an infinite number of world-sheet fields \cite{POL92}.
Whether deconfinement transition of gauge theories is related to the
Hagedorn transition of strings is also not quite clear.
The relation between the two, using 
AdS/CFT correspondence was discussed recently in 
\cite{KALYSATH} (see also earlier paper \cite{BKR})
 where it was found that in the strong coupling regime the 
deconfinement transition takes place before the Hagedorn transition.
The situation at weak coupling (the physical limit for gauge theories)
is not known.

Still, it is likely that the hot phases of gauge theories and
string theories have common physics. For example, using the fact that
in the deconfining phase  the free energy is proportional to $N^2$ while in
confining phase  it must be $O(1)$ one can argue \cite{THORN}  that
smooth world-sheets must disappear.  

All things considered, it  would  be very interesting 
to have a model where we know both, the
mechanism of confinement and the string description
at low energies on one side, and the detailed picture of deconfinement 
phase transition and the deconfining phase on  the other side. 
A model like this would be useful to learn as much as we can about the 
transition in string theory in a controlled setting.

Such a model in fact does exist. It is the confining Georgi-Glashow
model in $2+1$ dimensions.
It was shown by Polyakov long ago
\cite{POL75,POL77} that this theory is linearly confining 
due to the screening in the plasma of  monopole-instantons.
In 1996 Polyakov   proposed the description of the confining phase of
this theory  based on the so-called confining strings \cite{POL96}. 
Induced string action
can be explicitly derived for compact QED \cite{POL96}, 
\cite{confiningstringaction} and it was found that this nonlocal
action  has derivative  expansion describing  rigid string
\cite{Rigid1}. For more details
see \cite{confinigstringreferences} and references therein.

Recently the deconfining transition in this theory has been studied in
detail in \cite{KOVNER00}. The critical temperature for the 
transition has been calculated and it was shown that transition itself is in the 
universality class of the two dimensional Ising model.
The discussion in \cite{KOVNER00} was however purely field theoretical.
The purpose of the present paper is to relate the analysis of
\cite{KOVNER00} to Polyakov's confining string picture
and to discuss various aspects of the transition from the string perspective.
 
The outline of this paper is the following.
In Section 2 we discuss the properties of the confining string in 
the Georgi-Glashow model. Rather than introducing the Kalb - Ramond
field \cite{KALB} (like in \cite{POL96}) we use the direct 
correspondence between field configurations and the 
summation over closed string world-sheets. The integration over the 
field degrees of freedom leads then directly to the string action. 
This string will be referred to henceforth as the confining string.
An advantage of this derivation is that it makes explicit an important and
quite unusual property of the
confining string, namely that the fluctuations of this string with large 
momenta
(larger than the inverse thickness of the string) do not cost energy.
This is a dynamical extension of the so called zigzag symmetry introduced by
Polyakov \cite{POL96}.
We also show that the heavy charged particles of the Georgi-Glashow
model ($W^\pm$ bosons) appear as $D0$ branes in the string description.

In Section 3 we discuss the confining string at finite temperature.
First, we show that the Hagedorn temperature corresponds in the field theoretical
language to the temperature at which the monopole-instantons become 
bind in pairs. This temperature arises naturally in the Georgi-Glashow model, if one 
neglects the effects of the charged particles at finite temperature.
We also show that the world-sheet vortices which are usually
discussed in the context of the Hagedorn transition, physically correspond
to the trajectories of the 
endpoints of an open string propagating in the compact Euclidean time.
The actual mechanism of the deconfining transition 
in the Georgi-Glashow model though is quite different.
The transition is due to the appearance
of the plasma of charged particles, or in the string language - $D0$ branes.
Since a trajectory of a $D0$ brane bounds a world-sheet of an open string,
such a trajectory also corresponds to a vortex on the string world-sheet.
These vortices are however of a somewhat
different kind, since the string coordinates
satisfy the Dirichlet rather than Neuman boundary conditions.
At high temperature the string world-sheet is destroyed by appearance of 
these vortices, or in the target 
space picture the string is broken into short segments which connect the $D0$
branes. This mechanism, although it can be presented in the string language,
is essentially field theoretical in origin. It is driven
by physics on short distance scales - shorter than the physical thickness
of the string, where the stringy degrees of freedom are absent. Due to
the large thickness of the confining string and relative lightness of the
$D0$ branes,  the deconfining transition precedes the Hagedorn transition,
rendering the latter irrelevant.

Finally in Section 4 we conclude with discussion of our results.

\section{Confining strings in the Georgi-Glashow model.}

In this section we give the derivation of Polyakov's confining string and
discuss some of its striking physical properties. Note that in the present
derivation we do not use Kalb-Ramond fields. 

Consider the 2+1 dimensional $SU(2)$ gauge theory with
a scalar field $h$ in the adjoint representation. The action of the
theory is
\begin{equation}
S[A_\mu, h]= - \frac{1}{2 g^2} \int dx^3 \, {\rm tr}[F_{\mu \nu} \, F^{\mu \nu}] + \int dx^3
               \left( {\rm tr}(D_\mu h)^2 + 
\frac{\lambda}{4} (2{\rm tr} h^\dagger h -v^2)^2  \right)\,\,\, ,
\label{S0}
\end{equation}
where we use matrix notations $A_\mu = \frac{i}{2} A_\mu^a \tau^a $,  
$F_{\mu \nu} = \partial_\mu A_\nu - \partial_\nu A_\mu + [A_\mu,
A_\nu]$,
$h=\frac{i}{2} h^a \tau^a$, $D_\mu h = \partial_\mu h + [A_\mu, h]$ and 
the normalisation 
$tr(\tau^a \tau^b) = 2 \delta^{ab}$.
We will be working in the weakly coupled regime,  $v>>g^2$. 
In this regime the $SU(2)$
gauge group is broken to $U(1)$ by the large expectation value of the Higgs 
field. Perturbatively the photon 
associated with the unbroken group is massless, but the Higgs field as well as the 
$W^\pm$ gauge bosons are heavy with masses
$M^2_H = 2 \lambda v^2$ and $M^2_W = g^2 v^2$, respectively.
Furthermore, perturbatively the theory
looks like electrodynamics with spin one charged matter. 
At large distances the non-perturbative
effects become very important, so that the photon acquires mass and 
the $W^{\pm}$ become linearly confined
with a non-perturbatively small string tension. These non-perturbative effects 
according to Polyakov \cite{POL77}
are due to the contributions of monopole instantons.
The monopoles have Coulomb interactions and form a plasma 
in the sense of the 3-dimensional Euclidean path integral.
The ``Debye screening'' in this plasma provides for a finite photon mass.
The low energy physics of the theory is described by the effective Lagrangian
written in terms of the dual photon field $\phi$,
\begin{equation}
S[\phi]= \int dx^3 [\frac{g^2}{32 \pi^2} \, \partial_\mu \phi \partial^\mu \phi + 
        \frac{M^2_\gamma g^2}{16 \pi^2} \, (1-\cos \phi)] \,\,\, .
\label{S1}
\end{equation}
The monopole induced photon mass is $M^2_\gamma = \frac{16 \pi^2 \zeta}{g^2}$.
We use the notations of reference \cite{KOVNER00}, where the 
monopole fugacity is
$\zeta = \zeta_0 \frac{M^{7/2}_W}{g} \exp[-\frac{4 \pi M_W}{g^2} \epsilon(\frac{M_H}{M_W})]$,
with $\zeta_0$ a numerical constant while $\epsilon(\frac{M_H}{M_W})$ 
takes values between $1$ and $1.787$
and approaches unity for large values of ${\frac{M_H}{M_W}}$
(see \cite{KOVNER00} and references therein).

The effective action eq.(\ref{S1}) describes only the dynamics of the 
photon field and is valid at energies below the scale $M_W$.
The short distance physics will be important for us to some extent in the discussion
of the phase transition. We will come back to this question later, but 
for now let us discuss the effective Lagrangian eq.(\ref{S1}) as it
is. It exemplifies in a very simple manner the mechanism of confinement
in this theory. First, note that the theory has more than one vacuum state. In particular
$\phi=2\pi n$ for any integer $n$ is the classical ground state.
Therefore the classical equations of motion have wall-like solutions - 
where the two regions of space, say with $\phi=0$ and $\phi=2\pi$ are
separated by a domain wall. The action density per unit area of
such a domain wall is easily estimated from the action eq.(\ref{S1}) as
\begin{equation}
\sigma\propto g^2M_\gamma \,\,\, .
\end{equation}
As discussed in detail in \cite{POL75, POL77} 
the fundamental Wilson loop when inserted into 
the path integral with action
eq.(\ref{S1}) induces such a domain wall solution with the result
\begin{equation}
<W(C)>=\exp\{-\sigma S_m\}
\end{equation}
where $S_m$ is the minimal area subtended by the contour $C$.
Thus, the string tension is precisely equal to the wall tension of
the domain wall.
The domain wall in fact is nothing but the world-sheet of the 
confining QCD string.

The expression for the expectation value of the Wilson loop is the
starting point for Polyakov's derivation of the action for the confining
string in \cite{POL96}. We will, however, follow a slightly different
path which in our view makes it more straightforward to
understand some physical properties of the confining string.

\subsection{The string action.}
Since the string world-sheet is the domain wall in the effective action, we
will integrate in the partition function  all degrees of freedom
apart from those that mark the position of the domain wall.
To do this let us split the field $\phi$ into the continuous and the discrete
parts
\begin{equation}
\phi(x)=\hat\phi(x)+2\pi\eta(x)
\label{decomp}
\end{equation}
where the field $\hat\phi$ is continuous but bounded within the 
vicinity of one ``vacuum''
\begin{equation}
-\pi<\hat\phi(x)<\pi
\end{equation}
and the field $\eta$ is integer valued. Clearly, whenever $\eta(x)\ne 0$,
the field $\phi$ is necessarily not in the vicinity of 
the trivial vacuum $\phi=0$. Thus, the points in space where $\eta$ does
not vanish mark in a very real sense the location of a domain wall
between two adjacent vacua.

We now substitute the decomposition eq.(\ref{decomp}) into eq.(\ref{S1})
and integrate over $\hat\phi$. The partition function is
\begin{equation}
Z=\int {\cal{D}}\hat\phi {\cal{D}}\eta \exp(-
\int dx^3 [\frac{g^2}{32 \pi^2} \, 
\partial_\mu(\hat \phi+2\pi\eta) \partial^\mu (\hat\phi+2\pi\eta) + 
\frac{M^2_\gamma g^2}{16 \pi^2} \,(1- \cos \hat \phi)] \,\,\, .
\label{Z2}
\end{equation}
At weak coupling the field $\hat\phi$ can be integrated out in the 
steepest descent approximation. This means that the equations of motion
for $\hat\phi$ have to be solved in the presence of the external source
$\eta$ and keeping in mind that $\hat\phi$ only takes values 
between $-\pi$ and $\pi$. Let us assume for this solution that
the surfaces along which $\eta$ does not vanish are few and far 
between. Also we will limit the 
possible values of $\eta$ to $0,1,-1$. In the context of the
effective Lagrangian eq.(\ref{S1}) this is the ``dilute gas''
approximation. However as a matter of fact, 
as we will see below, this exhausts all physically allowed
values of $\eta$. 
Then, the qualitative structure of the solution for $\hat \phi$ is
clear. 
In the bulk, where $\eta$ vanishes, the field $\hat\phi$ satisfies
normal classical equations. When crossing a surface $S$ on
which $\eta=1$, the field 
$\hat\phi$ must jump by $2\pi$ in order to cancel the contribution
of $\eta$ to the kinetic term, since otherwise
the action is UV divergent. 
Thus the solution is that of a ``broken wall'' - far from the surface
the field $\hat\phi$ approaches its vacuum value $\hat\phi=0$, it raises
to $\pi$ when approaching $S_m$, then jumps to $-\pi$
on the other side of the surface and again approaches zero far from the 
surface. The profile of the solution for smooth $S$ (in this 
case a plane) is depicted on Fig.1. 

\begin{center}
\epsfig{file=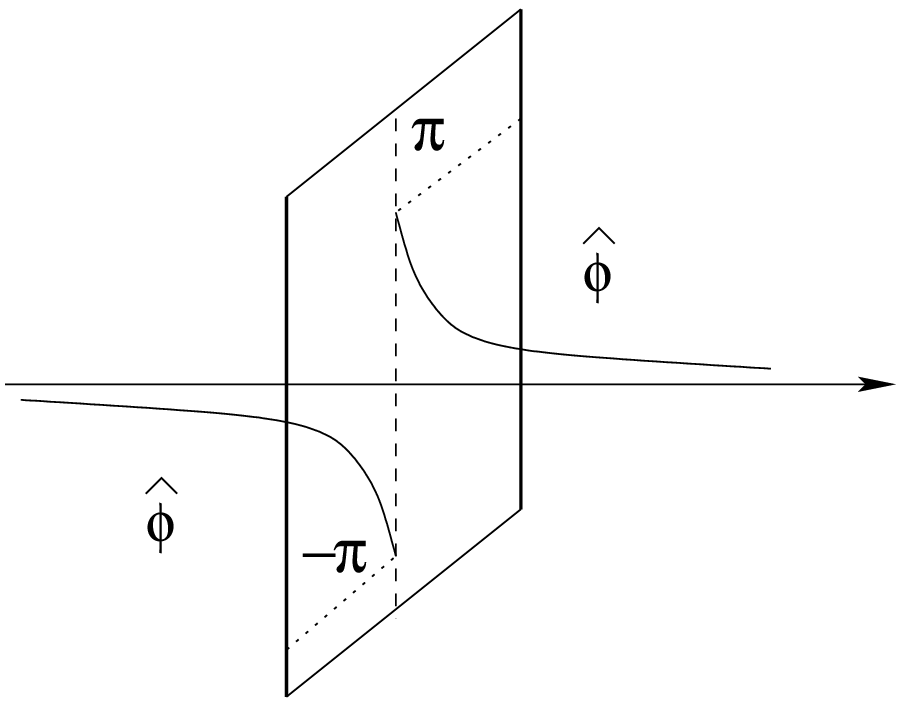, width=9.cm}
\end{center}
\baselineskip=13pt
\centerline{\small{Figure 1: Schematic representation of the 
solution profile of}}
\centerline{\small{the $\hat\phi$ field as a function of the coordinate
perpendicular to the domain wall.}} 

\vspace{1.cm}

\baselineskip=20pt plus 1pt minus 1pt

Since outside the surface
the field $\hat\phi$ solves classical equations of motion, clearly the 
profile of $\hat\phi$ is precisely the profile of the domain wall
we discussed above. The only difference is that this wall is broken
along the surface $S_m$ and the two halfs of the solution are displaced
by $2\pi$ with respect to each other. The sole purpose of
this discontinuity, as noted
above is to cancel the ``would be'' UV divergent contribution
of $\eta$ to the kinetic term in the action. Thus the action
of our solution is precisely
\begin{equation}
S[\eta]=\sigma S=\sigma 
\int d^2 \xi \frac{\partial X_\mu}{\partial\chi^\alpha} 
\frac{\partial X_\mu }{\partial\xi_\alpha} \,\,\, .
\end{equation}
where $\xi$ are the coordinates on the surface $S$ and $X_\mu$ are
the coordinates in the 3-dimensional space-time.
This is precisely the Nambu-Gotto action of a free string, which classically
is equivalent to the well known Polyakov´s string action. 

Of course the confining strings in the Georgi-Glashow model are not free.
The thickness of the region in Fig.1 in which the field $\hat\phi$
is different from zero, is clearly of order of the inverse
photon mass $M_\gamma^{-1}$. Thus, when two surfaces come within the 
distance of this order they start to interact. In principle, this
interaction can be calculated by just finding the classical solution
for $\hat\phi$ in this situation. 
Now, however we want to discuss one particular
property of the confining strings - their rigidity. 

\subsection{The string is soft ... And therefore rigid!}

The Nambu-Gotto action we have derived in the previous subsection
is of course only the long wavelength approximation to the actual action of
the confining string. It is only valid for string world-sheets which
are smooth on the scale of the inverse photon mass.
Expansion in powers of derivatives can be in principle performed
and it will give corrections to the action on scales comparable 
to $M^{-1}_\gamma$.
However, for
our confining string strange things happen in the ultraviolet.
Physically the situation is quite peculiar, since the action of the string
has absolutely no sensitivity to the changes of the world-sheet on short 
distance scales. This should be obvious from our derivation of the string 
action.
Suppose that rather than taking $\eta=1$ on an absolutely smooth surface, we
make the surface look the same on the scale $M_\gamma^{-1}$ but add to it some
wiggles on a much shorter distance scale $d$, as in Fig.2. 

To calculate the action
we have now to solve the classical equation for the field $\hat\phi$ with
the new boundary condition - the surface of the discontinuity is wiggly. 
This new boundary condition 
changes the profile of the classical solution only within the thickness $d$
of the old surface. However, since the action of the
classical solution is in no way dominated by the region of space close 
to the discontinuity, the action of the new solution will be the same
as that of the old solution with the accuracy $dM_\gamma$.
Thus, all string configurations which differ from each other only
on small resolution scales $d\ll M^{-1}_\gamma$ have with this accuracy 
the same energy!. The string is therefore extremely soft, in the sense that it
tolerates any number of wiggles on short distance scale without cost in energy,
Fig.3. In particular, since the 
string tension for our string is much greater than
the square of the photon mass, $\sigma/M_\gamma^2=g^2/M_\gamma\gg 1$, 
fluctuations on the scale of the string tension are not 
penalised at all.

\begin{center}
\epsfig{file=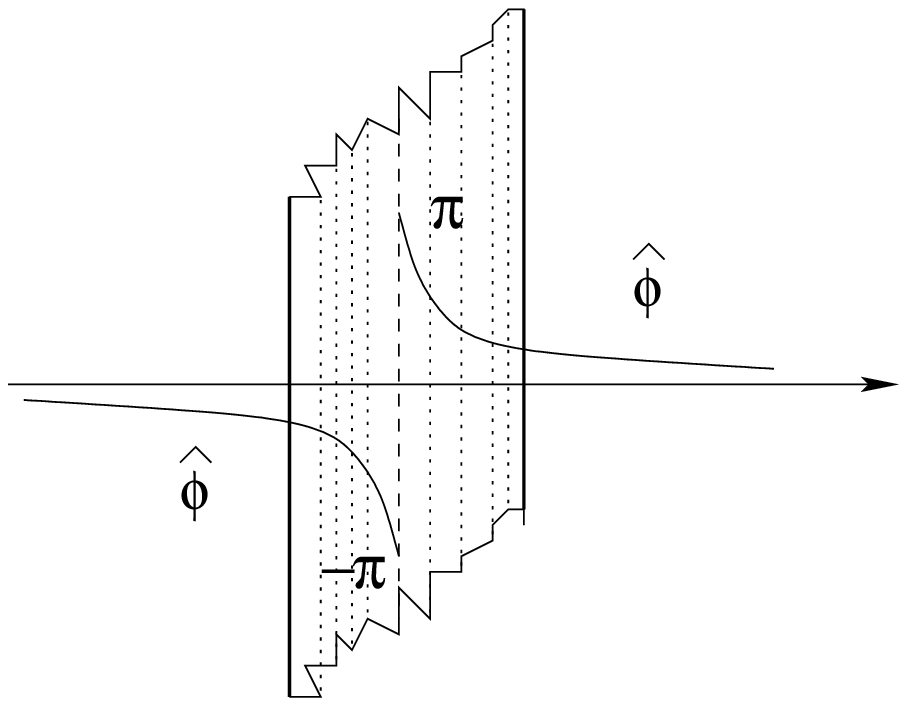, width=9.cm}
\end{center}
\baselineskip=13pt
\centerline{\small{Figure 2: Schematic representation 
of the solution profile of}}
\centerline{\small{the $\hat\phi$ field with a ``rough'' domain wall profile.}} 

\vspace{1.cm}

\baselineskip=20pt plus 1pt minus 1pt

We believe that this independence of the action on short wave length 
fluctuations is a dynamical manifestation of the so called zigzag symmetry
introduced by Polyakov\cite{POL96}. Indeed, Polyakov notes
that if a segment of a string goes back on itself, physically 
nothing has happened and so such a zigzag should not cost any action.
This situation is an extreme example of a wiggle we have just discussed - 
a wiggle with infinite momentum. What happens physically, is that
not only an infinite momentum modes, but also finite but large momentum
string modes do not cost energy.

The confining string is therefore very different from a weakly
interacting string usually considered in the string theory.
In the weakly interacting string momentum modes with momentum of
order of the square root of the string tension carry energy
which is of the same order. In the confining string on the other hand,
these momentum modes do not carry energy at all. Thus, we do not expect
the spectrum of the confining string to contain states with
large spatial momentum (heavy string states with low angular
momentum)\footnote{Indeed, the states with high angular momentum may very well be absent
too, due to the fact that a long string can decay into a $W^+-W^-$ pair.
Therefore, we believe that most of the string states will not appear
in the spectrum of the Georgi - Glashow model contrary to belief expressed
in \cite{POL96}.}.

\baselineskip=20pt plus 1pt minus 1pt

\begin{center}
\epsfig{file=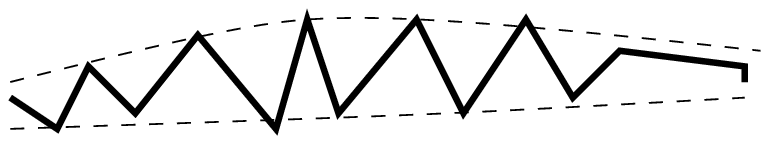, width=7.cm}
\end{center}
\baselineskip=13pt
\centerline{\small{Figure 3: The solid line represents the actual {\it weak} 
string. The}} 
\centerline{\small{dashed lines mean the contours of the effective thick 
string. }} 

\vspace{1.cm}

\baselineskip=20pt plus 1pt minus 1pt

In a somewhat perverse way, this softness of the ``mathematical'' string 
leads to rigidity of the physical string.
The point is the following. As it is obvious from the previous discussion, the
high momentum modes are in a sense ``gauge'' degrees of freedom. The
action of any string configuration does not depend on them.
Consider a calculation of some physical quantity $O$ in the string
path integral. If $O$ itself does not depend on the string high momentum
modes, the integration over these modes factors out. If on the other hand
$O$ does depend on them, then its average vanishes since their fluctuation
in the path integral is completely random.
This is absolutely analogous to the situation in gauge theories,
where all observables must be gauge invariant, and for calculation of
those the integration over the gauge modes always factors out, independently
of the observable under consideration.

Thus, for all practical purposes, we should just exclude the high
momentum string modes from the consideration altogether 
(``gauge fix'' them to zero). 
For example when calculating the entropy of our string we should only take 
into account the states which are different on the course grained level
with the course graining scale of order $M^{-1}_\gamma$.
This means that 
our string is intrinsically ``thick''. If we still want to describe
this situation in terms of a thin string with the string tension
$\sigma$, we must make the string rigid so that any bend on the scale
smaller than $M_\gamma^{-1}$ be suppressed.
Thus, such a string theory must have a curvature term with the coefficient
of order $\sigma/M^2_\gamma$

 The derivative expansion for the
confining string action has been discussed in the literature 
(see \cite{confinigstringreferences}) and
references therein). It has in general the form
\begin{equation}
 S = \int d^2 \xi \sqrt{g} g^{ab} D_a \vec{x} \left( \sigma + s D^{2} +
..\right) D_b \vec{x} \,\,\, ,
\end{equation}
where $g$ and $D_a$ are the determinant and the covariant derivative with
respect to induced  metric $ g_{ab} = \partial_a \vec{x} \partial_b
\vec{x}$ on the embedded world-sheet $\vec{x}(\xi_1, \xi_2)$.
The rigidity is controlled by the second term. In the derivative expansion
the stiffness $s$ is negative, and the system is stabilised by
the higher order terms.
Our argument shows that the cumulative effect of (the infinite number of) the
 higher order terms at short distances ($D\le M_\gamma$)
is equivalent to a large and positive curvature term
\begin{equation}
 S = \int d^2 \xi \sqrt{g} g^{ab} D_a \vec{x} \left( \sigma +k
{\sigma\over M_\gamma^2} D^{2}\right) D_b \vec{x}
\end{equation}
with $k$ a coefficient of order one.
Hence in this way the fact that the confining string lacks real stringy 
high momentum degrees
of freedom translates itself into large rigidity term in the effective 
description.
We emphasise again that the rigidity is very large, and is effective 
on the distance scales much larger {\it parametrically} than the ``natural'' 
string scale 
given by the string tension.

\subsection{The $W^\pm$ bosons - the $D0$ branes of confining string.}

As discussed in \cite{KOVNER00} 
the deconfining phase transition in the Georgi-Glashow model
is driven by the $W^\pm$ bosons. It is therefore important
to understand how they fit into the string picture.

The Polyakov effective Lagrangian eq.(\ref{S1}) is only 
valid at energies much lower than $M_W$. 
This in principle does not preclude us from discussing $W^\pm$ in its 
framework, since being charged particles they have a long range-low momentum
field component associated with them, and this ``Coulomb'' field is 
in principle describable
in the framework of eq.(\ref{S1}). We will not be able to describe the 
internal structure of $W^\pm$ on the scale of their Compton wave length,
but this is of no importance to us in any case.
There is however one important element that we have to
add to eq.(\ref{S1}) before the discussion of $W^\pm$ can proceed.

As mentioned in the introduction, the Georgi-Glashow model
possesses the global magnetic $Z_2$ symmetry. This symmetry in the confining
phase is spontaneously broken in the vacuum \cite{thooft1, KOVNERZN}.
How is this $Z_2$ symmetry represented in the effective
Lagrangian eq.(\ref{S1})? The answer is, that in fact 
the field $\phi$ does not take values in $R$, but is rather a phase. 
More precisely,
the field $\chi$, where
\begin{equation}
\phi=2\chi
\end{equation}
is the phase field which takes values between $0$ and $2\pi$.
One can define a vortex operator\cite{KOVNERZN}
\begin{equation}
V=({g^2\over 8\pi^2})^{1/2}\exp\{i\chi\} \,\,\, .
\end{equation}
In terms of this operator the Lagrangian eq.({\ref{S1}) is
\begin{equation}
S=\int dx^3 [  \partial_\mu V \, \partial^\mu V^* 
       - \frac{M^2_\gamma }{4} \, (V^2+V^{*2})] \,\,\, .
\label{efft}
\end{equation}
The magnetic $Z_2$ symmetry acts as
$V\rightarrow -V$.

Since the field $\chi$ is a phase, it is obvious that the effective theory
eq.(\ref{efft}) allows topologically nontrivial configurations
with non-vanishing winding of $\chi$. This is precisely how
the $W^\pm$ bosons are represented in this effective theory.
According to \cite{thooft1,KOVNERZN}, the electric current is identified with the 
topological current in eq.(\ref{efft}),
\begin{equation}
{\frac{g}{\pi }}J_{\mu }=i\epsilon _{\mu \nu \lambda }\partial _{\nu
}(V^{\ast }\partial _{\lambda }V) \,\,\, .
\end{equation}
Thus, charged states carry unit winding number of $\chi$. The $W^+$ ($W^-$) 
boson is a state with positive (negative) unit winding of $\chi$.
The action also has to be augmented by a higher derivative term in order
that the winding configuration has the ``core'' energy equal
to $M_W$. By core energy we mean the energy concentrated at distances of order
of the UV cutoff of our effective theory which is not contained in the
low momentum field $\chi$. This higher derivative term is
\begin{equation}
\delta S=\int d^3x {1\over g^4 M_W} (\epsilon_{\mu\nu\lambda}\partial_\nu V^*
\partial_\lambda V)^2
\end{equation}
The density of this action does not vanish only at the points where the 
phase of
$V$ is singular - that it the points of winding. For a closed curve $C$
of length $L$
which carries the winding, the contribution of this extra
term to the action is $M_WL$, which is precisely the action
of the world-line of a massive particle of mass $M_W$.

To represent a $W$ boson in the string language, let us consider a
path integral in the presence of one such particle. To create a winding state
with world-line $C$ we should insert a source in the path integral which
forces a winding on the field $\chi$. In terms of the field $\phi$
that would mean that it has to change by a $4\pi$ when going around $C$.
The relevant partition function is
\begin{equation}
Z[C]=\int D\phi\exp\{-\int dx^3 [\frac{g^2}{32 \pi^2} \, (\partial_\mu \phi-
2\pi j_\mu(x)) (\partial^\mu \phi-2\pi j_\mu(x)) + 
        \frac{M^2_\gamma g^2}{16 \pi^2} \, (1-\cos \phi)] -M_WL\}\,\,\, .
\end{equation}
Here
the ``external current'' $j_\mu$ is 
\begin{equation}
j_\mu(x)=n^1_\mu(x)\delta(x\in S_1)+n^2_\mu(x)\delta(x\in S_2)
\label{surf}
\end{equation}
where $S_1$ and $S_2$ are two surfaces which both terminate on the curve $C$, 
and the unit vectors $n^1$ and $n^2$ are the normal vectors to these surfaces.
The shape of the surfaces on which the current $j_\mu$ does not
vanish is illustrated on Fig.4.

\begin{center}
\epsfig{file=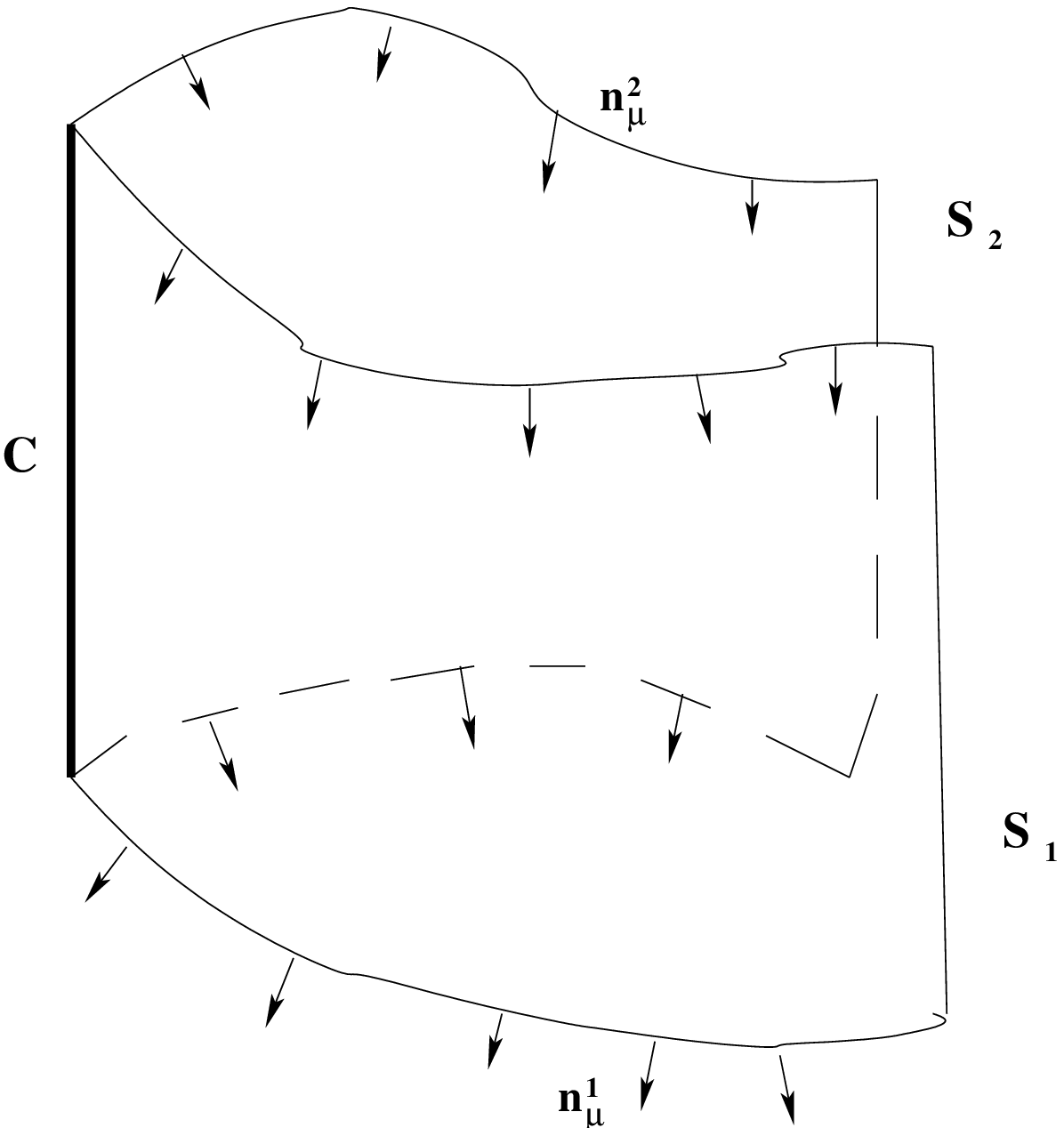, width=9.cm}
\end{center}
\baselineskip=13pt
\centerline{\small{Figure 4: The two surfaces $S_1$ and $S_2$ along which the current $j_\mu$ is non-vanishing.}}
\centerline{\small{ The surfaces come together at the
line $C$ which is the world-line of the $W$ boson.}}
\vspace{1.cm}

\baselineskip=20pt plus 1pt minus 1pt
The insertion of $j_\mu$ forces the field $\phi$ to jump across the surface
$S_1$ by $2\pi$ and again jump by $2\pi$ across $S_2$ in order to cancel
the, otherwise UV divergent contribution of $j_\mu$. Thus when going around
$C$ the field $\phi$ changes by $4\pi$ and therefore $C$ is the world-line of 
the $W^+$ boson \footnote{We note that the position 
of the surfaces $S_1$ and $S_2$
is arbitrary as long as they both terminate on $C$. In particular they could
coincide, which is the form used in \cite{KOVNER00}. Here we prefer to use
a more general form with noncoinsiding surfaces for reasons which will become
clear immediately.}.

As before, splitting the field variable $\phi$ into $\hat\phi$
and $\eta$ we see that the presence of $j_\mu$ just shifts the variable $\eta$
by one on the two surfaces $S_1$ and $S_2$. The integration over $\hat\phi$
at fixed $\eta$ is performed in exactly the same way as before. The only
difference now is that for any given $\eta$ one has two extra string 
world-sheets
along $S_1$ and $S_2$\footnote{We again stress that after the integration
over $\eta$ the result will not depend on exact position of $S_1$ and $S_2$.
However at fixed $\eta$ the two surfaces introduced in eq.(\ref{surf})
specify the positions of the two extra world-sheets.}. 
We thus see that the field theoretical path integral in the presence 
of a $W^+$ boson, in the string representation is given by the sum
over surfaces in the presence of a $D0$ - brane, which serves as a 
source for two extra string world-sheets. 
Usually $D0$ branes are thought of as infinitely heavy. The situation
in the GG model is very similar in this respect. They are not infinitely heavy,
but very heavy indeed since the mass of $W$ is large on all scales relevant
to zero temperature physics.
The contribution of
the $D0$ brane to the partition function is suppressed by a very small factor
$\exp\{-M_WL\}$, and vanishes for infinitely large system.
As we will see in the next section however, the situation changes dramatically
at finite temperature, where one dimension of the system has finite extent.

Incidentally, going back to our definition of the vortex field $V$, we see that
in between the two world-sheets the value of $V$ is negative, while outside
them it is positive. Thus we have created domain of the second vacuum of
$V$ in between $S_1$ and $S_2$. It may be easier to visualise the situation
with both $W^+$ and $W^-$ present. In this case the surfaces $S_1$ and $S_2$
terminate on one side on the world-line of $W^+$, and on the other side on the
world-line of $W^-$ see Fig.5. They are thus 
boundaries of a closed domain of the 
second vacuum of the field $V$. In the infrared therefore our 
strings are nothing but the Ising domain walls, and the pair of $D0$-branes
creates an Ising domain.

Note, that in physical terms there are only two
distinct vacua in the model $<V>=1$ and $<V>=-1$. Thus having two domain walls
is the same as having a wall and an anti-wall, and if they coincide spatially
such a configuration is equivalent to the vacuum. A configuration of
$\eta$ with $\eta=2$ on a closed surface is physically equivalent
to vacuum. Therefore the values that $\eta$ is allowed to take
are limited to $0$,$1$ and $-1$. 

\begin{center}
\epsfig{file=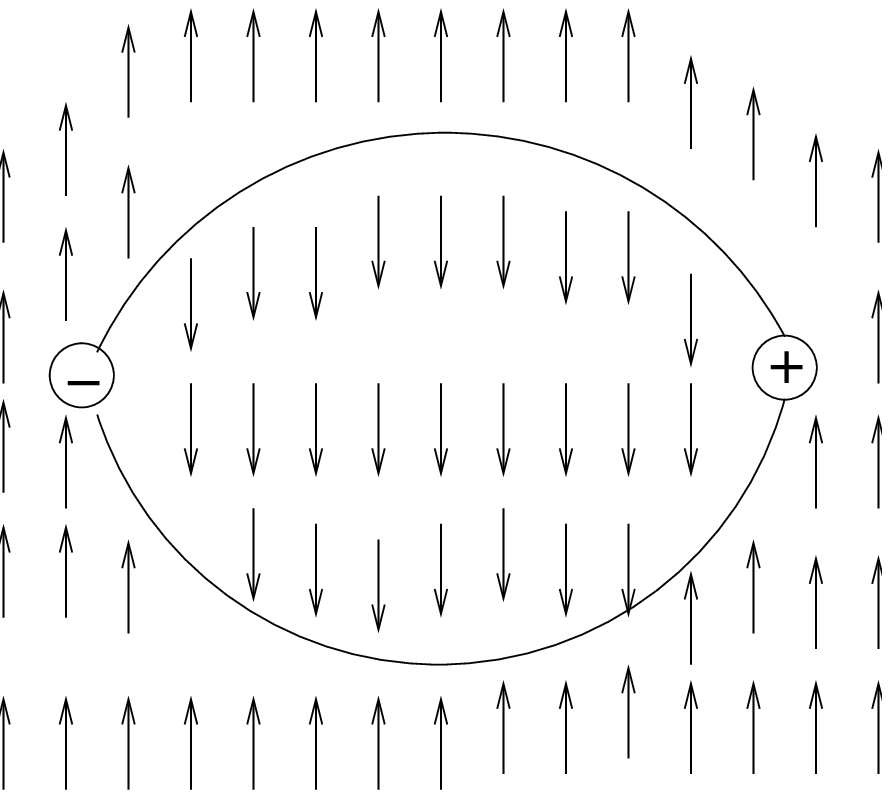, width=6.cm}
\end{center}
\baselineskip=13pt
\centerline{\small{Figure 5: The Ising domain created by the two strings
connecting the positions of $W^+$ and $W^-$ particles.}}
\centerline{\small{For static particles the configuration propagates in the time direction, perpendicular to the plane of the page.}}
\vspace{1.cm}

\baselineskip=20pt plus 1pt minus 1pt
We close this section by noting, that the reason we have two string
world-sheets terminating on a $D0$ brane is that the field theory in question
has only fields in adjoint representation of $SU(2)$. 
We can imagine adding heavy fundamental particles
to the model. We would then have also allowed configurations of one string 
world-sheet terminating on a $D0$ brane. These $D0$ branes would however 
be physically different and would have a different mass and therefore a 
different weight in the path integral. In the field theoretic terms, presence 
of the fundamental charges changes drastically the properties of the $Z_2$
magnetic symmetry, turning it into a local rather than a global symmetry 
\cite{fosco}.

\section{How does the string melt?}
Now that we have an understanding of the basic physical properties of the 
confining string we can move on to the analysis of the deconfining
phase transition. The question we want to address is what is the mechanism
of the melting of the string in the deconfined phase. 
In this respect the basic mechanism that has been discussed in the framework
of the string theory is the Hagedorn transition, where the limiting
temperature is reached due to excitations of the high energy spectrum of 
the string. It has also an interpretation as the BKT transition due to
the condensation of the vortices on the string world-sheet.

In this section we will first explain how this
mechanism is  related
to the well understood physics of deconfinement
in the GG model. It turns out that the Hagedorn transition
corresponds to ``confinement'' or binding 
of magnetic monopoles at high temperature.
Then we show that the vortex on the world-sheet 
should be understood as the world-line 
of the end point of an open string. 
Thus the ``condensation of vortices on the world-sheet'' is
the statement that once the monopoles are bound, and 
there is no linear potential between charges any longer, the 
thermal ensemble contains  arbitrarily long strings. If open strings 
exist in the theory as dynamical objects, they
appear in the thermal ensemble (world-sheet vortices). 
If only closed strings are allowed, as is the case in the
GG
model, the thermal ensemble is dominated by
closed strings of arbitrary length.
World lines of charged particles also map into world-sheet vortices
in the string description, although these vortices are slightly
different.
The presence of arbitrarily long strings would mean 
deconfinement of charges, or ``plasma phase''  also of these vortices.
 
Although such a mechanism of transition is a logical possibility,
it turns out that 
the actual mechanism in the Georgi-Glashow model
is completely different. Due to the thickness of the physical string,
the transition happens long before the ``Hagedorn temperature''. It occurs 
because the density of charged particles ($D0$ branes) becomes large
enough, so that the distance between them becomes smaller than the width of
the string. So the thermal ensemble is dominated by 
configurations in which the length of the strings is 
shorter than their thickness. At this point of course the
string picture becomes useless. The theory really becomes the plasma
of weakly interacting $D0$-branes.
This destruction of the string ``from within'' 
is quite peculiar from the string
point of view.  In a sense it is of distinctly field theoretical nature,
rather than a string theoretical nature. It occurs due
to physics on the scale smaller than the string thickness, where as we 
have seen the string language is inadequate, since 
no string modes exist at these momenta.

\subsection{Deconfinement in the Georgi-Glashow  model.}

We start with a brief recap of the main points of \cite{KOVNER00}.
Within the GG model there are two mechanism for the deconfining
transition that one can contemplate.

The first one is the binding of magnetic monopoles at high temperature.
This mechanism was first suggested in \cite{AZ}.
The point here is the following. 
In the Euclidean path integral formalism the 
monopoles form Coulomb gas with the interaction 
``potential'' decreasing 
as $1/r$ at large distances.
However, at finite temperature the interaction in the infrared
 becomes logarithmic. 
The reason is that the finite temperature path integral is formulated
with periodic boundary conditions in the Euclidean time
direction. The field lines
are therefore prevented from crossing the
boundary in this direction. The magnetic  
field lines emanating from the monopole have
to bend close to the boundary and go parallel to it. So, effectively
the whole magnetic flux is squeezed into two dimensions. 
The length of the time direction is $\beta=1/T$, and thus clearly the
field profile is two dimensional on distance scales larger than
$\beta$. Two monopoles separated by a distance larger than $\beta$
therefore interact via a two dimensional rather than a three dimensional
Coulomb potential, which is logarithmic at large distances. Since the
density of the monopoles is tiny $\rho_M\propto M^2_\gamma$, already at
extremely low temperatures $T\propto M_\gamma$ the monopole gas
becomes two dimensional.
The strength of the logarithmic interaction is easily calculated.
The magnetic flux of the monopole far enough from the core 
spread evenly in the
compact direction.
The field strength should have only components parallel to the spatial
directions. Since the
total flux of the monopole is $2\pi/g$, the field
strength far from the core is 
$\tilde F_i={T\over g}{x_i\over x^2}$,
and thus the strength of the infrared logarithmic interaction
is $T^2/g^2$.

Two dimensional Coulomb gas 
is known to undergo the BKT phase transition. Due to the peculiar
property that for the monopoles the strength of the logarithmic interaction
grows with temperature, the high temperature phase corresponds to the 
monopole binding phase.
The transition temperature above which the monopoles are bound is
\begin{equation}
T_{MB}={g^2\over 2\pi} \,\,\, .
\label{MB}
\end{equation}
Below this temperature the photon should be massive, while above this
temperature it should be massless since the cosine term in the
Lagrangian eq.(\ref{S1}) is irrelevant.
Thus one may expect the deconfining transition at $T=g^2/2\pi$ which is
in the universality class of 2D XY model.

The other candidate mechanism is the appearance of the charged plasma
of $W^\pm$.
If one neglects the monopole effects and considers a ``non-compact'' theory,
the potential between charged particles is logarithmic. At finite
temperature one has a certain, albeit very small density of $W$'s. 
Thus the system again resembles a two dimensional Coulomb gas.
This gas undergoes a BKT transition from a confined phase to a plasma phase
at $T_{NC}=g^2/8\pi$. If this where the mechanism of the transition,
the universality class would be again that of 2D $XY$ model.

The truth however lies somewhere in between \cite{KOVNER00}.
It turns out that at $T_{NC}$ the density of $W$'s is very small, so that
the transition does not occur since the charges are bound by the linear
potential. However at $T_{C}=g^2/4\pi$ the mean distance between
$W^\pm$ in the ensemble becomes equal to the thickness of the confining
string. At this point it does not make sense any more to
think of the thermal ensemble as dilute gas of $W^\pm$ pairs bound by  
strings, but the ensemble rather
looks like a neutral plasma. Indeed it is shown rigorously in \cite{KOVNER00}
that the transition occurs at $T_C$. The universality class of
the transition is that of the 2D Ising model 
corresponding to the restoration of 
the magnetic $Z_2$ symmetry. The transition 
can be pictured as condensation of 
Ising domains which fill the interior of $W^+$-$W^-$ bound states.

\subsection{The monopole binding as the Hagedorn transition.}

Let us now consider the phase transition from the string perspective.
First, in analogy with the discussion in the previous subsection, let
us completely disregard the charged particles. From the string
point of view this means that we neglect possible contributions of the
heavy $D0$ branes, and thus are entirely within the theory of closed strings.
Naively one expects in such a theory existence of Hagedorn temperature,
beyond which the 
string can not exist. In an almost free string this temperature
is of order of the string scale.
One can visualise this phenomenon in simple terms.
Consider a closed string of a given fixed length $L$.
Let us calculate the free energy of such a string. The energy of the string is
\begin{equation}
E=\sigma L
\end{equation}
The number of states for a closed string of the length $L$ scales exponentially
with $L$
\begin{equation}
N(L)=\exp\{\alpha L\}
\end{equation}
The dimensional constant $\alpha$ is determined by the physical thickness 
of the string. Imagine that the string can take only positions allowed
on a lattice with the lattice spacing $a$. Then clearly the number of possible
states is $z^aL$, where $z$ is the number of order one, equal to
the number of nearest neighbours on the lattice\footnote{We disregard
in this argument the fact that the string has to close on itself. This extra condition would lead to a prefactor with power 
dependence on $L$. Such a prefactor is not essential for our argument, 
and we therefore do not worry about it.}. 
In this simple situation $\alpha=a\ln z$.
The only natural
lattice spacing for such a discretization is the thickness of the string.
For an almost free string the thickness is naturally the same as the
scale associated with the string tension. Thus the entropy
is
\begin{equation}
x\sqrt\sigma L \,\,\, ,
\end{equation}
with $x$ a number of order one.
The free energy then is
\begin{equation}
F[L]=\sigma L-xT\sqrt\sigma L \,\,\, .
\end{equation}
At the temperature 
\begin{equation}
T_H={1\over x}\sqrt\sigma
\end{equation}
the free energy becomes negative, which means that strings of arbitrary length
appear in the thermal ensemble in a completely unsuppressed way.
The thermal vacuum becomes a ``soup'' of arbitrarily long strings.
Thus, effectively the ``temperature dependent'' string tension vanishes
and it is not possible to talk about strings anymore in the hot phase.
 For more details about this ``random walk'' description of hot
strings see \cite{hotsoup} and references therein.

The situation is very similar to the BKT phase transition, where the free 
energy of a vortex becomes negative at the critical temperature, and the
vortices populate the vacuum in the hot phase.

In the string partition function language this is just a
restatement of the well known 
fact that the partition function diverges in the 
sector with the topology of a torus which winds around the compact
Euclidean time direction.
Fixing the unit winding in the Euclidean time physically corresponds to
the calculation of the free energy in the sector with one closed string.
The integration over all possible lengths of the string is the cause of the
divergence of the partition function at high temperature.

The same physical effect is there for the confining GG string.
There is one difference, which however turns out to be crucial
for the nature of the phase transition. The 
thickness of the GG string is not given by the string 
tension. Rather it is equal to the inverse photon mass, and thus
\begin{equation}
\alpha\propto M_\gamma\propto {\sigma\over g^2}\ \ \ .
\end{equation}
We thus have for the confining GG string
\begin{equation}
T_H={\sigma\over \alpha}\propto g^2
\end{equation}
This is indeed the correct magnitude of the critical temperature
as discussed in the previous subsection. 

The noteworthy feature of this formula is, that
the Hagedorn temperature of the confining string
is much higher than the string scale. This is easy to understand
because the entropy of the thick string is much smaller than that 
of the free string due to the fact that high momentum modes of the confining
string do not contribute to the entropy at all, 
as discussed in the previous section.
Thus, one needs to heat the string to a much higher temperature for entropy
effects to become important.

Since we have completely neglected the possibility of
appearance of $D0$ branes (charged particles), 
the transition we have been discussing is the string 
representation of the monopole binding transition in the Georgi Glashow model.
At the point where the monopoles bind, the photon of the GG model becomes
massless and thus the string tension disappears and the thickness of the 
string diverges. In the Hagedorn picture this is just the dual statement 
that the ensemble is dominated by the
infinitely long strings.
The BKT nature of both transitions underlines this point.

\subsection{Vortices on the world-sheet - open strings and charged particles.}

Sometimes the Hagedorn transition is discussed in terms of
the vortices on the world-sheet. What is the physical nature of these
objects? 

Consider a string world-sheet with the topology of a sphere with a
vortex-anti-vortex pair. When going around the vortex location
on the world-sheet, the compact coordinate $x_0$ varies from $0$ to $\beta$. Since
this is true for any contour of arbitrarily small radius, which encircles the
vortex, this means that physically the location of the vortex in fact
corresponds in the target space not to a point but rather to a line
which winds around the compact direction. The vortex-anti-vortex pair on 
the world-sheet thus represents an {\it open} string which winds around the compact
direction. Fig.6 illustrates how the open string world-sheet, equivalent to
a cylinder is transformed by a conformal transformation into a sphere with
two singular points - the vortex-anti-vortex pair.

If the string theory in question does have an open string sector, 
the configurations with arbitrary number of vortex-anti-vortex pairs
contribute to the finite temperature partition function.
The Hagedorn transition then can be discussed in this sector rather than in the
closed string sector. Not surprisingly,
the discussion is exactly the same as in
the previous subsection. The fact that we now have an open rather than closed 
string does not change the entropy versus energy argument. One finds then that
at $T<T_H$ the vortices on the world-sheet are bound in pairs. 
The corresponding target space picture is that the "ends" of the open strings
are bound by linear potential and therefore long open strings do not contribute
to the thermal ensemble. In exactly the same way, long closed strings are also 
absent
from the ensemble. At $T>T_H$ the vortices unbind and appear in the ensemble
as the Coulomb gas. Thus a typical configuration contains lots of open
strings (as well as lots of arbitrarily long closed strings) since the 
energy of such strings is overwhelmed by the entropy.

\begin{center}
\epsfig{file=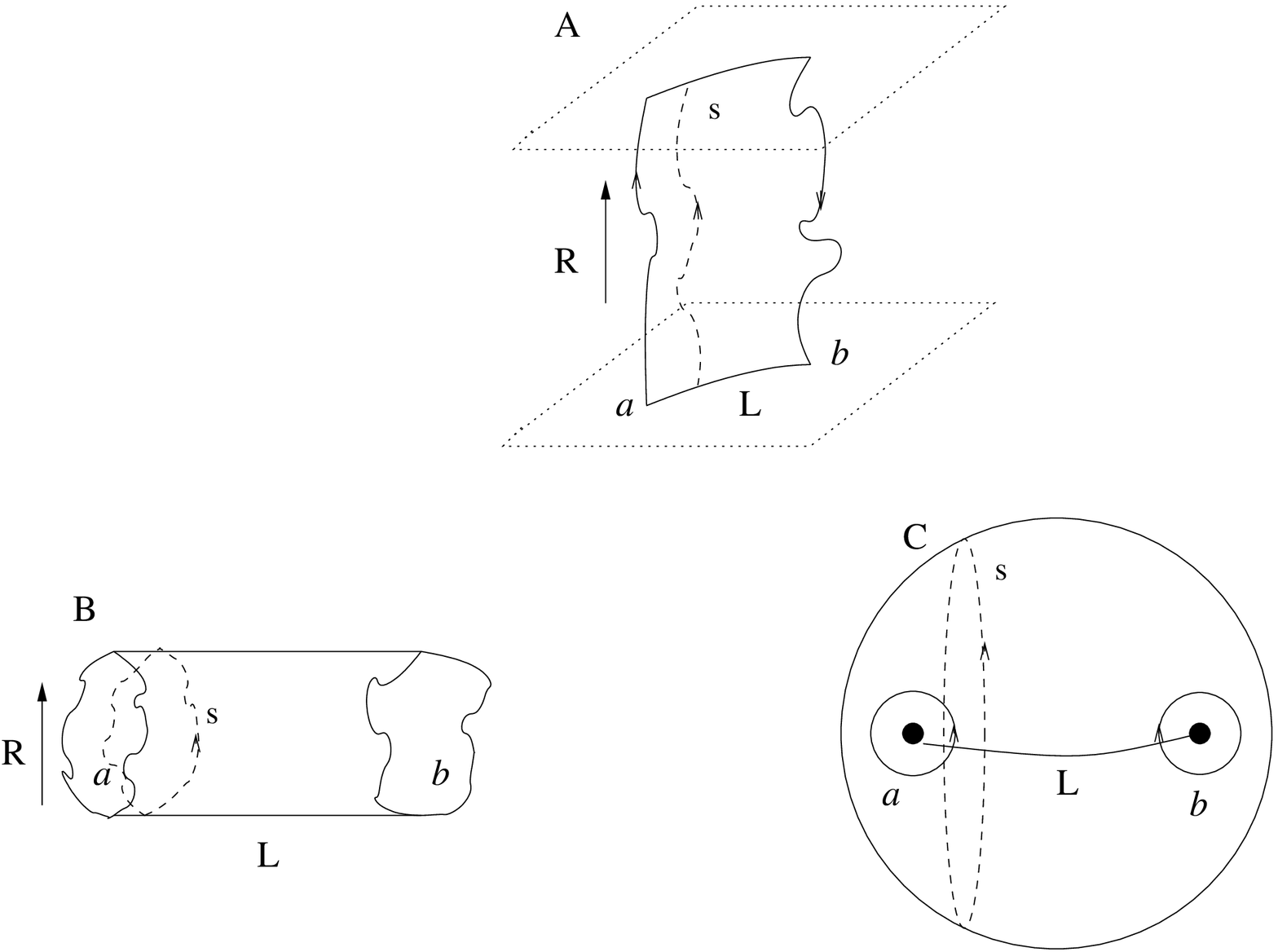, width=12.cm}
\end{center}
\baselineskip=13pt
\centerline{\small{Figure 6: The open string world-sheet}}
\centerline{\small{conformally  transformed into
a closed string world-sheet with a vortex-anti-vortex pair.}} 
\centerline{\small{L is the length of the string, $a$ and $b$ are
locations of the vortex and the anti-vortex}}
\centerline{\small{on the world-sheet and $R$ represents
the compact dimension.}}
\vspace{1.cm}

\baselineskip=20pt plus 1pt minus 1pt

Note however, that the existence of the transition 
in this context
is entirely independent of 
the presence or absence of the vortices. As we saw above, the transition 
can be understood
purely on the level of the closed string. It is driven by the string
fluctuations.
Thus, even if the theory does not have an open string sector, 
the transition is still there. For example there
are no open strings in the  Georgi Glashow model. Nevertheless,
if we neglect the effects of $W^\pm$ the Hagedorn transition is still
there and it coincides with the "monopole binding" transition.

The actual phase transition in the Georgi-Glashow model is however
not driven by monopole binding. In the string language nevertheless, 
it is also due to the proliferation of vortices on the world-sheet. Those 
are however vortices of a somewhat different type.
The same conformal transformation that turned the open string boundary 
into a point can be used to turn into a point the world-line of
a $D0$ brane. As discussed above, a (fundamentally) charged particle
which couples to the GG confining string is indeed a $D0$ brane.
Thus a string world-sheet representation of a pair of particles with opposite
charge is also a vortex-anti-vortex pair on a sphere\footnote{We note 
that the name ``vortex'' is particularly 
apt in this case, since 
the same charged particle, which generates the vortex on the world-sheet, 
is also a vortex of the dual photon field $\phi$ in the Polyakov effective 
theory eq.(\ref{S1}) as discussed in the previous section.}.

There are some important differences between these vortices and the ones that
represent the ends of the open string. 

First, for an open string the
non-compact "spatial" coordinates satisfy Neuman boundary conditions. Thus even
as $x^0$ winds, the other coordinates $x^i$ can take arbitrary values 
close to the vortex. One can see this from the action
\begin{equation}
S = \sigma \int d^2 \xi \partial_a x^0 \partial_a x^0 + 
\sigma \int d^2 \xi \partial_a x^i \partial_a x^i
\end{equation}
where the dynamics of $x^i$ sector is absolutely unrelated to the
dynamics of $x^{0}$. Thus even though a vortex in $x^0$ sector  can  be
considered as a boundary in the target space,  the boundary
conditions for the components $x^i$ are free. 
In other words there is no boundary action 
induced by vortices in a theory of closed strings. 
For a very heavy $D0$ brane on the other hand the boundary conditions
are of the Dirichlet type. 
Thus $x_i$'s are constant close enough to the vortex location.
If the
mass of a $D0$ brane is finite, 
there is a nontrivial boundary action describing a
massive particle. In that case one has  non-conformal boundary
conditions compatible with the finite mass of the $D0$ brane \cite{KW}.

Another difference, is that since the $D0$ brane is an independent degree 
of freedom, in principle its mass is a free parameter. Thus the fugacity
of the $D0$ brane vortex is an independent parameter, and
the contribution of such a vortex to the thermal ensemble
is determined by this fugacity.
The thermal physics may 
depend on this extra parameter.

To get back to the theory at hand, the Georgi-Glashow confining string has
neither an open string sector nor $D0$ 
branes which are sources of a single string.
The dynamical objects which couple to the string 
are the charged $W^\pm$ particles, which
have an adjoint charge and therefore are sources of a pair of strings.

\subsection{$D0$ branes destroy the string.}
 
Each $D0$ brane has two strings emanating from it. Thus a pair of
branes propagating in compact imaginary time is conformally equivalent to a
pair of ``double vortices'' as in Fig.7.
The singular points are still vortices as before, since going around
such point one travels once around the compact time direction. But
now two string world-sheets
are permanently glued together at the location of a vortex.
The fact that two world-sheets are glued at the location of the vortex
is the manifestation of the $Z_2$ magnetic symmetry of the GG model. The
region of space between the two world-sheets is separated from the
region of space on the outside, reflecting the fact that it
constitutes a domain of a different vacuum.

\begin{center}
\epsfig{file=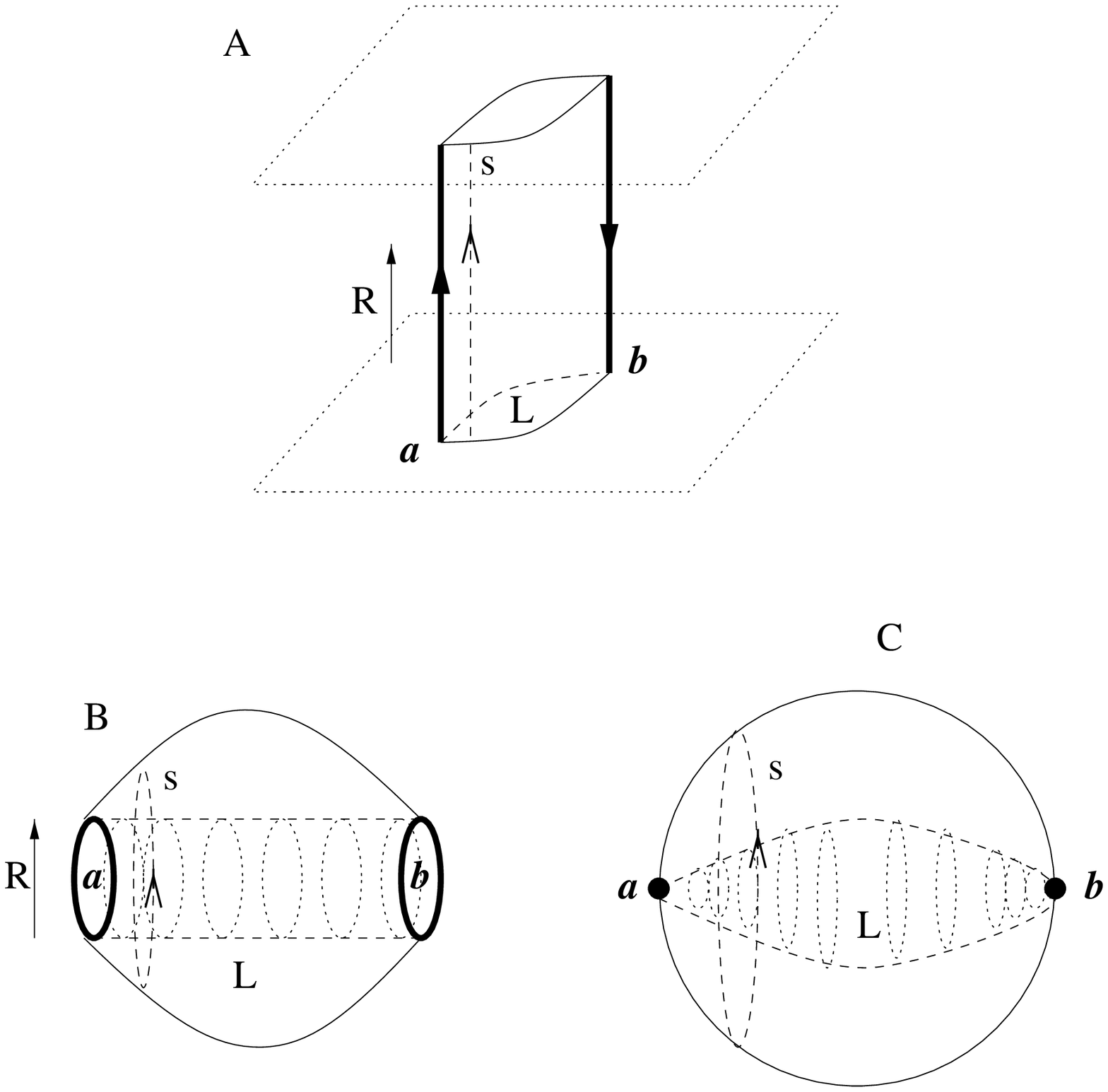, width=12.cm}
\end{center}
\baselineskip=13pt
\centerline{\small{Figure 7: The D0 brane-anti-brane ($W^+$ - $W^-$) pair 
with two strings connecting them}} 
\centerline{\small{conformally transformed into a string world-sheet 
with two ``double vortices''.}}

\vspace{1.cm}

\baselineskip=20pt plus 1pt minus 1pt

At low temperatures such configurations in the thermal ensemble are
rare. But at the critical temperature their density becomes so large
that the ensemble is dominated by multi-vortex configurations. In the
string picture those are configurations with multiple points of
``gluing''. Importantly the $D0$ branes are not just pairwise connected by two
strings,
but rather form a network where all of them are connected to each
others. Clearly entropy wise such configurations are much more
favourable, and there is also no loss of energy when the distance
between the $D0$ branes is of the order of the string thickness.
As an example we show in Fig.8 a configuration
with four mutually connected $D0$ branes.

\begin{center}
\epsfig{file=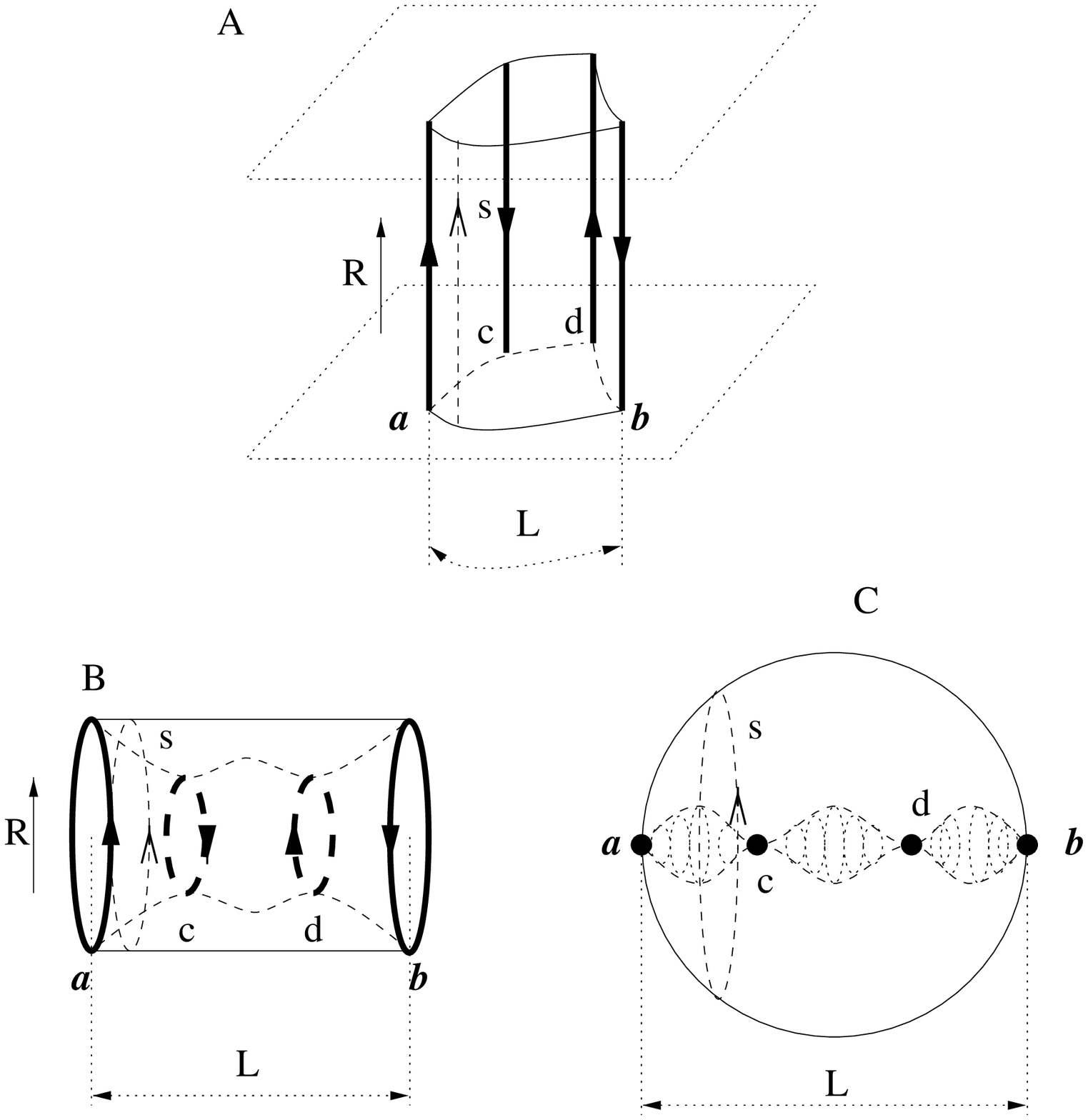, width=12.cm}
\end{center}
\baselineskip=13pt
\centerline{\small{Figure 8: Four branes connected by strings.}}
\centerline{\small{Such
configurations become important above the transition.}}

\vspace{1.cm}

\baselineskip=20pt plus 1pt minus 1pt

This string-based description of the transition has to be taken with a
grain of salt. The relevant physics takes place on short distance
scales -
of the order of the thickness of the string. On these scales, as we
explained above, the string modes are practically absent. Thus even
though we have showed the string segments on Fig.8, these segments are
so short
that there is no string tension associated with them. Thus the
mechanism of the transition is essentially field theoretical rather
than string theoretical.

To understand this point better consider in a little more detail the 
thermal ensemble of $W^\pm$. The crucial point is that at distance
scales $d\ll M_\gamma^{-1}$ the interaction between them is Coulomb
rather than linear. The gas of charges with Coulomb interaction
has itself a transition into plasma phase. This transition has nothing to do
with the long range linear interaction and occurs at the temperature $T_{NC}$
which is
four times smaller than the Hagedorn temperature as discussed in
subsection 3.1 \cite{KOVNER00}.
At this temperature $W^\pm$ become "free" in the sense that they cease to
care about the Coulomb part of the potential. The crucial question
however is how large is the density of $W^\pm$ at this point.
If the density at $T_{NC}$ where large and the average distance between
$W$'s where smaller than $M_\gamma^{-1}$, the transition would actually
occur at $T_{NC}$, since the  long range linear 
part of the potential would be entirely irrelevant. As it happens, in the GG model
this is not the case, and the density of $W$'s is small. Thus at $T_{NC}$
there is a certain rearrangement of the thermal ensemble on short distance scale, 
but at long distances nothing happens - the string still confines. 
However, at $T_C=T_H/2$ the density of $W$
reaches the critical value and the transition occurs. Note that at this temperature
the large length fluctuations of the string are still suppressed - we are far below
the Hagedorn temperature. The string is destroyed not due to "stringy" physics
of the Hagedorn transition, but due to the short distance field theoretical
effects: the fact that fugacity of $W$ is relatively large and that the interaction
at short distances is Coulomb and not linear.

\section{Discussion.}

Let us summarise the discussion of the two previous sections.
The confining string in the Georgi-Glashow model is thick. Its physical
thickness is much larger than the square root of the inverse string tension.
Physics on the distance scales smaller than the thickness of the string is
essentially field theoretical, since the high momentum string degrees 
do not contribute. The heavy $W$-bosons behave like $D0$ branes - they are sources
of a pair of confining strings.

At high temperature the appearance of the $W$ - bosons, $D0$ branes in the 
thermal ensemble is understood in the world-sheet picture as appearance
of vortices and anti-vortices on the world-sheet. The density of these
vortices is governed by the fugacity of the $W$ - bosons. When the 
distance between the branes becomes equal to the thickness of the string,
the transition occurs. The thermal ensemble above the transition
is dominated by states with large numbers of $D0$ branes, connected
 to each other
by short segments of the string.
Talking about segments of string in these circumstances is really only a
mnemonic,
since the confining string does not have any modes at momenta corresponding to 
the inverse length of those strings. Thus, the string world-sheet disappears.
The string disintegrates into $D0$ branes.
The transition has nothing to do with the Hagedorn transition,
which corresponds to appearance of arbitrarily long strings (closed or open).
Thus the deconfining transition occurs at the temperature lower than the 
Hagedorn temperature of the confining string.

One lesson that we learn from this, is that the UV structure of the string
is extremely important. In this particular case, the ultraviolet sector
contains $D0$ branes, and at high enough temperature they dominate physics,
thus superseding the stringy transition mechanism. 

How universal is this situation?. Does deconfinement 
always precede Hagedorn?. It is hard to tell. Naively one could think,
that since the mass of the $D0$ brane is a free parameter, it can be
made arbitrarily large. Thus, one could have a situation that at the
Hagedorn temperature the density of $D0$ branes is still small, and 
the string remains intact all the way up to $T_H$. If that were the case, the 
Hagedorn transition would indeed be realized. However, within the GG model
the mass of the $W$ boson is not free, once the string tension and the width
of the string are fixed. Making $W$ heavier also makes the string 
thicker, and this conspiracy always leads to marginalization of the
Hagedorn transition, at least as long as the coupling is weak.
In the strong coupling limit (light $W$) things may be different. 
Strong coupling corresponds to the
pure Yang Mills theory. In this case, at least at large $N$ the spectrum
is believed to be much closer to a stringy spectrum\cite{TEPER}. So
perhaps the Hagedorn transition takes over. On the other hand, even at
large $N$ the confining string has a finite thickness. The analog of
the $W$ mass - half the mass of a glueball - is also finite. 
Thus, even in the large $N$ limit there is no parameter
which would tell us that the $D0$ brane mechanism is irrelevant.

In fact, universality arguments suggest that the $D0$ brane
condensation mechanism prevails all the way to  strong coupling. The
transition in the $SU(2)$ gauge theory is supposedly in the Ising
universality class, just like in the Georgi-Glashow model. On the
other hand the Hagedorn transition is the BKT transition and is thus 
in the XY universality class. It seems therefore that at least in the
$SU(2)$ case the deconfining transition is not of the Hagedorn type.
For larger $N$ one can show that the transition in the weakly coupled
limit is again second order and is not in the universality class of
$U(1)$\cite{inprep}. The Hagedorn transition is again of the $U(1)$ type.
By the usual universality prejudice, we expect the weak and strong
coupling regimes to be in the same universality class. If that is the case,
here the Hagedorn transition does not win either.
The situation however can be different in other dimensionalities, 
and our analysis has nothing to say about that.

The explicit solution of the Georgi-Glashow model does
provide interesting examples of phenomena discussed in the framework
of the string theory. We saw how the open strings and $D0$ branes
realize the vortices on the world-sheet. The $D0$ branes we dealt with
have two strings attached to them. In $SU(N)$ theories
each $D0$ brane would have $N$ strings, and the corresponding vortex
would have $N$ string world-sheets glued at its location.
We have understood the connection between the Hagedorn transition and
the binding
of magnetic monopoles. Finally, we have seen how the string is
destroyed by the condensation of $D0$ branes.

We hope that the simple but nontrivial physics discussed here will 
eventually be of help in the attempts to understand hot string theory.

\section*{Acknowledgements}

The work of I.K. was supported by PPARC Grant PPA/G/O/1998/00567. 
The research of A.K. was supported by PPARC.
M.S. has been supported by CONICET of Argentina, Fundaci\'on 
Antorchas of Argentina and The British Council.

\vfill

\end{document}